\documentclass{article}
\usepackage{spconf,amsmath,graphicx}

\title{Modeling Intrapersonal and Interpersonal Influences for Automatic Estimation of Therapist Empathy in Counseling Conversation}

\name{Dehua Tao$^1$, Tan Lee$^1$, Harold Chui$^2$, Sarah Luk$^2$}
\address{
  $^1$ Department of Electronic Engineering \quad 
  $^2$ Department of Educational Psychology\\The Chinese University of Hong Kong}

\begin{document}
% \ninept
%
\maketitle
%
% About 100 to 150 words;
\begin{abstract}
Counseling is usually conducted through spoken conversation between a therapist and a client. The empathy level of therapist is a key indicator of outcomes. Presuming that therapist's empathy expression is shaped by their past behavior and their perception of the client's behavior, we propose a model to estimate the therapist empathy by considering both intrapersonal and interpersonal influences. These dynamic influences are captured by applying an attention mechanism to the therapist turn and the historical turns of both therapist and client. Our findings suggest that the integration of dynamic influences enhances empathy level estimation. The influence-derived embedding should constitute a minor portion of the target turn representation for optimal empathy estimation. The client's turns (interpersonal influence) slightly surpass the therapist's own turns (intrapersonal influence) in empathy estimation effectiveness. It is noted that concentrating exclusively on recent historical turns can significantly impact the estimation of therapist empathy.

\end{abstract}
\begin{keywords}
counseling conversation, therapist empathy, intrapersonal influence, interpersonal influence, attention mechanism
\end{keywords}
\section{Introduction}
\label{sec:intro}

Counseling is a common therapeutic practice in psychology. It is typically conducted as a verbal conversation between a therapist and a client, with the primary goal of providing a supportive environment for the client to express emotion freely, alleviate distress, and navigate challenges in life. In the field of psychotherapy, empathy is defined as ``the therapist's sensitive ability and willingness to understand the client's thoughts, feelings, and struggles from the client's point of view" \cite{rogers1995way}. The level of empathy demonstrated by the therapist plays a pivotal role in the counseling process. It is regarded as a key indicator of psychotherapy outcome and therapeutic effectiveness \cite{elliott2011empathy, moyers2013low, elliott2018therapist}.

% self and inter-speaker
% The behavioral expressions of participants in a conversation are not solely driven by their internal mental states but are also influenced by the responses of their counterparts
Conversation is an interactive activity. Participants engage in communication through verbal and non-verbal behaviors. The behavioral expressions are influenced by the participants' own internal processes as well as the responses of their counterparts \cite{schmidt2008dynamics}. Such intrapersonal and interpersonal influences have been examined in conversation-related studies, e.g., emotion recognition in conversations \cite{hazarika2018conversational,hazarika2018icon,yeh2019interaction}. In the context of counseling, prior research \cite{soma2020coregulation} revealed that both therapist and client exhibited intrapersonal and interpersonal patterns of emotional arousal throughout the conversation. In \cite{paz2021intrapersonal}, the intrapersonal and interpersonal vocal affect dynamics within and between clients and therapists were investigated. It showed a significant correlation between these dynamics and the outcomes of psychotherapy. As such, it is plausible to infer that the therapist's expression of empathy is influenced not only by his/her own past behavioral states but also by his/her perception of the client's behavior.
% As such, it is plausible to infer that the therapist's expression of empathy at a given moment is influenced by both their internal state at that moment and the residual impact of previous conversational interactions.

% Therapist empathy is usually measured over the entire counseling conversation. That is, a human observer goes through the conversation and assigns an overall score of empathy level. In a long conversation, the therapist may express high empathy in certain parts and low empathy in the other parts, depending on the degree of engagement and the content of conversation. 
In a typical counseling conversation, the therapist and the client take turns to speak. A speaker turn is defined as a time period during which only one person speaks. The conversation can be viewed as a sequence of speaker turns, each turn being spoken by either the therapist or the client. In the present study, an influence model is proposed for estimating the empathy level of each therapist turn. An attention mechanism is employed to quantify intrapersonal and interpersonal influences on each therapist turn. Subsequently, these turn-level estimations are integrated to produce an overall rating of empathy for the whole conversation.
% the estimated turn-level empathy scores are integrated to produce an overall rating of empathy for the whole conversation.

For the modeling of intrapersonal and interpersonal influences, we focus on the vocal behaviors of both therapist and client. Each speaker turn is represented by the acoustic properties of speech within the turn \cite{xiao2013modeling,xiao2014modeling,imel2014association,xiao2015analyzing}. Experimental results indicate that the inclusion of both intrapersonal and interpersonal influences enhances the estimation of the therapist's empathy level. By examining the weighted combination of influence-derived embedding and the acoustic representation of therapist turn, it is found that the influence embedding should constitute a relatively small fraction of this combination to attain optimal empathy estimation. The interpersonal influence derived from the client's turns marginally outperforms the intrapersonal influence from the therapist's own turns in empathy level estimation. Additionally, it is observed that focusing solely on recent historical turns can have a substantial influence on therapist empathy estimation.
% By examining the weighted combination of influence-derived embedding and the acoustic representation of the therapist's turns, 

\section{Dataset}
\label{sec:dataset}

A speech corpus of counseling conversations is used in this research. The corpus, named CUEMPATHY, contains $156$ audio recordings of conversations \cite{tao2022cuempathy}. The conversations involve $39$ distinct therapist-client dyads. That is, each therapist and each client appear in only one dyad. For each therapist-client dyad, $4$ conversations are included in the corpus. The recordings were collected during counseling practicums for therapist trainees at the Chinese University of Hong Kong. The participating clients were adults seeking psychological assistance on a wide range of concerns, including stress, emotions, relationships, and personal growth. All therapists and clients spoke Hong Kong Cantonese. Each conversation was about $50$ minutes long. The study was approved by the institutional review board, and informed consent was obtained from all participating therapists and clients.

% The TES is a nine-item observer rating scale to assess affective, cognitive, attitudinal, and attunement aspects of therapist empathy.
For each of the $156$ conversations, the therapist's empathy was subjectively rated according to the Therapist Empathy Scale (TES) \cite{decker2014development} by trained observers. TES is a nine-item rating scale that covers various aspects of therapist empathy, including affective, cognitive, attitudinal, and attunement dimensions. A score for each item is given on a 7-point scale from $1=\textit{not at all}$ to $7=\textit{extremely}$ after observers complete watching a videotaped counseling session. Thus the total TES score for a conversation ranges from $9$ to $63$, with a higher score indicating a higher level of therapist empathy. To evaluate the inter-rater reliability, about $40\%$ of the conversations ($62$ conversations) were rated independently by two observers. The intra-class coefficient was $0.90$, indicating excellent inter-rater agreement \cite{cicchetti1994guidelines}.

A total of $118$ conversations are selected from the $156$ conversations to form $2$ subsets with polarized empathy scores in our experiments. The first subset consists of $61$ conversations of empathy scores from $42$ to $56.5$, with a mean score of $46.34\pm3.58$. These conversations are labeled as the high-empathy category. The second subset contains $57$ conversations with empathy scores from $18$ to $36$, with a mean of $30.40\pm4.79$. They are labeled as the low-empathy category. Across the $118$ conversations, there are $39$ distinct therapists involved. Among them, $17$ therapists are categorized as having both high and low empathy, $12$ therapists are classified solely under the high-empathy category, and $10$ therapists are exclusively classified under the low-empathy category. Table \ref{tab:data_info} summarizes the speech data used in our experiments. Given a counseling conversation, our goal is to classify it as either belonging to the high or low-empathy category using the speech of both the therapist and the client.
% our goal is to classify the therapist as exhibiting either a high or low empathy level using the speech of both the therapist and the client.
% Across the $118$ conversations analyzed, there are $39$ distinct therapists involved. Among these therapists, $17$ engage in conversations that are categorized as both high and low empathy. Additionally, $12$ therapists have conversations exclusively classified under the high-empathy category, while $10$ therapists have conversations solely classified under the low-empathy category. 

\begin{table}[htb]
\caption{Summary of counseling conversations used in this study. Average speech time per conversation (AvgTime), average number of speaker turns per conversation (AvgTurn), and average duration per turn (AvgDur) are calculated for each speaker.}
\label{tab:data_info}
\centering
\resizebox{1.0\linewidth}{!}{
\begin{tabular}{c|c|c|c}
Speaker       & AvgTime (min) & AvgTurn & AvgDur (sec) \\ \hline\hline
Therapist (T) & 14.89                                                                                & 139                                                                       & 6.03                                                                           \\ \hline
Client (C)    & 33.66                                                                                & 138                                                                      & 12.93                                                                           \\
\end{tabular}}
\end{table}

\section{The Proposed Method}
\label{sec:method}

Consider a counseling conversation that contains $N$ speaker turns, which are represented as $\mathcal{C}=(x_1^C,x_2^T,x_3^C,x_4^T...,x_N^{\phi})$. The turns, denoted by $x_i^{\phi}$, alternate between client (C) and therapist (T) in chronological order, where $i\in[1,N]$ and $\phi\in\{C,T\}$. The empathy level for $\mathcal{C}$ is expressed as a binary variable, with $1$ indicating high empathy and $0$ signifying low empathy. The therapist empathy is determined in two steps: (1) applying an attention-based influence model (AIM) to estimate the probability for each therapist turn to be high-empathy; (2) aggregating the turn-level estimated probabilities by median fusion to determine the overall empathy level, high or low, for the whole conversation.
% Given that only the overall empathy level of the therapist throughout the conversation is available, with individual turn empathy levels being unspecified, we propose a two-step process to ascertain the conversation-level therapist empathy. First, an attention-based influence model (AIM) is applied to estimate the probability for each therapist turn to be high-empathy. Second, the turn-level estimated probabilities are aggregated by median fusion to determine the overall empathy level, high or low, for the whole conversation.

\subsection{Attention-based Influence Model (AIM)}
\label{ssec:iiam}
\begin{figure}[htb]
  \centering
  \includegraphics[width=0.75\linewidth]{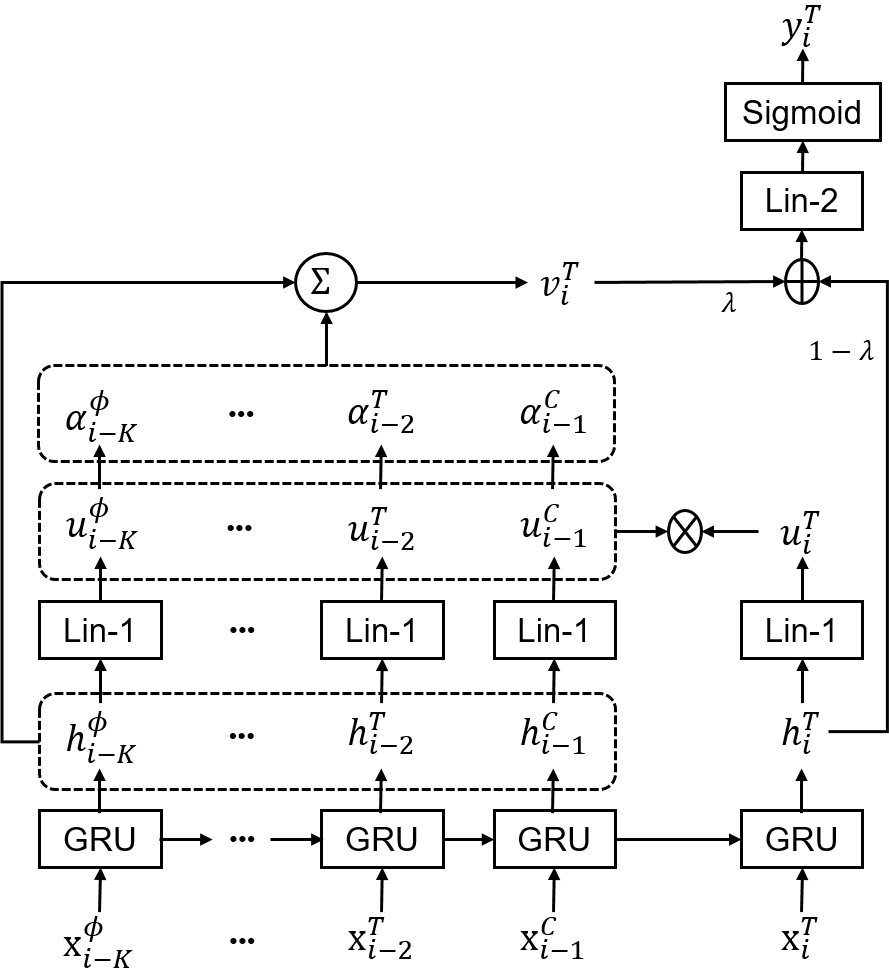}
  \caption{Illustration of the AIM.}
  \label{fig:intermodel}
\end{figure}

% The estimation of high empathy
The estimation of empathy level for the therapist turn $x_i^T$ is done by considering the previous turns from both client and therapist. To model such intrapersonal and interpersonal influences, an influence window is defined for the target turn $x_i^T$. The window covers $K$ historical turns, denoted as $w_i^T=(x_{i-K}^{\phi},..,x_{i-2}^T,x_{i-1}^C)$. $K$ is referred to as the size of influence window. The structure of AIM is illustrated in Figure~\ref{fig:intermodel}.

\noindent
\textbf{Turn encoder layer:} An unidirectional Gated Recurrent Unit (GRU) \cite{cho-etal-2014-learning} is adopted to read the speaker turns in the influence window $w_i^T$ as well as the target turn $x_i^T$ to model the sequential relationship between them. The GRU output of the speaker turn $x_i$ is obtained as $h_i = \text{GRU}(x_i, h_{i-1})$, by which the historical information from previous turns is incorporated.

% The GRU output of each turn $x_i$ is then through a linear layer with the tanh activation function to obtain the turn representation $u_i$. The encoding process is expressed by Eq. ((\ref{eq:1})).

% \begin{equation}
% \begin{aligned} 
%   &h_i = \text{GRU}(x_i, h_{i-1}) \\
%   &u_i = \text{tanh}(W_xh_i + b_x)
%   \label{eq:1}
% \end{aligned}
% \end{equation}

\noindent
\textbf{Attention-based influence layer:} The GRU output $h_i$ is fed into a linear layer with the tanh activation function to generate the turn-level representation, formulated as $u_i=\text{tanh}(W_xh_i)$. The $u_i^T$ for the target turn serves as a query to determine the importance weights assigned to all turns in the influence window. These weights quantify both intrapersonal influence from the therapist's own turns and interpersonal influence from the client's turns. Specifically in Eq. (\ref{eq:2}), the dot-product attention with softmax function is implemented between the query $u_i^T$ and the representation $u_{i-k}$ of each turn in the influence window. The weighted sum of GRU outputs for the $K$ historical turns is calculated to give the influence embedding $v_i^T$ for the target turn. The embedding $v_i^T$ encapsulates comprehensive dynamic influences on the therapist's expression of empathy at the target turn.

\begin{equation}
\begin{aligned}
  &\alpha_{i-k} = \frac{\text{exp}(u^\top_{i-k}u_i^T)}{\sum_{k=1}^K \text{exp}(u^\top_{i-k}u_i^T)}, k\in[1,K] \\
  &v_i^T = \sum_{k=1}^K \alpha_{i-k}h_{i-k}
  \label{eq:2}
\end{aligned}
\end{equation}

\noindent
\textbf{Output layer:} The refined representation for the target turn is obtained by combining the GRU output $h_i^T$ and the influence embedding $v_i^T$, expressed as $\bar{h}_i^T=(1-\lambda)h_i^T+\lambda v_i^T$. The parameter $\lambda$ is referred to as the influence scale. It determines the extent to which the dynamic influences are retained in estimating the empathy level. To represent the probability of the target turn exhibiting a high level of empathy, a linear layer followed by a sigmoid function is applied to map the refined representation to a value between $0$ and $1$. The probability is calculated as $y_i^T=\text{sigmoid}(W_o\bar{h}_i^T+b_o)$.

\subsection{Fusion Layer}
\label{ssec:fusion}

Following the computation of $y_i^T$ for each therapist turn in the conversation $\mathcal{C}$, the overall empathy level of therapist in the conversation is determined by fusing turn-level estimates. The method of median fusion is adopted, expressed as $y^{est}=\text{median}(..., y_{i-2}^T,y_i^T,y_{i+2}^T,... )$, where $y^{est}\in[0,1]$. This $y^{est}$ represents the probability of the conversation $\mathcal{C}$ being classified as high-empathy. While other fusion methods can be applied, conversation-level fusion is not the main focus of this research.

During the training phase, the objective function is defined as the binary cross entropy between the target $y^{tgt}\in\{0,1\}$ and the predicted probability $y^{est}$, as detailed in Eq. (\ref{eq:3}). In the inference phase, the given conversation $\mathcal{C}$ is classified as high-empathy if the probability $y^{est}$ exceeds $0.5$, and conversely as low-empathy if $y^{est}$ falls below $0.5$.

\vspace{-0.2cm}
\begin{equation}
\begin{aligned}
  &\mathcal{L}=-\frac{1}{L}\sum_{l=1}^L[y^{tgt}\text{log}y^{est}+(1-y^{tgt})\text{log}(1-y^{est})]
  \label{eq:3}
\end{aligned}
\end{equation}
where $L$ is the number of conversations in a mini-batch.

\section{Experimental Setup}
\label{sec:expset}

A 6-fold cross-validation (CV) is performed on the $118$ counseling conversations. In each iteration, conversations from $4$ folds are utilized for training, a single fold is designated for development, and another single fold is reserved for testing. Given that the number of conversations in the high and low-empathy categories is balanced, the model performance is evaluated using the metric of binary classification accuracy.

\subsection{Model Configuration}
\label{subsec:modelconf}

% The eGeMAPS contains acoustic parameters that can be used to measure human perception, which has been extensively used in, such as emotion recognition \cite{eyben2015geneva}, speech intelligibility prediction \cite{Xue2019}, and detection of Alzheimer’s dementia \cite{haider2019assessment}. 
The acoustic properties of a speaker turn are encoded by the 88-dimensional eGeMAPS feature vector \cite{eyben2015geneva}. The turn-level feature vector is computed by the openSMILE toolkit \cite{eyben2010opensmile} with the default script. The speaker-dependent z-normalization is performed for each dimension of turn-level features. The hidden size of GRU is set to $64$. The size of the linear layer (Lin-1) is set to $32$. For training, a batch size of $8$ is utilized, and an Adam optimizer with $\beta_1=0.9$ and $\beta_2=0.999$ is employed, along with a learning rate of $0.001$. The model is trained on a fixed number of $100$ epochs. The optimal model is determined based on the classification accuracy on the development data. The mean of classification accuracies on the 6-fold CV is used to indicate the model's overall performance.

\subsection{Baseline Models}
\label{subsec:baseline}

To assess the effectiveness of the proposed AIM in modeling the intrapersonal and interpersonal influences when estimating the empathy level of therapist turn, four baseline models are explored in the experiments. The conversation-level median fusion applied in the baseline models is identical to that used in the AIM.

\noindent
\textbf{IM}: This model feeds the GRU output of target turn to the output layer without implementing any attention mechanism.

\noindent
% \textbf{AIM\_\text{$\phi$}}, where \text{$\phi\in\{C,T\}$}
\textbf{AIM\_T} or \textbf{AIM\_C}: In this model, the influence window includes exclusively the turns of either the therapist (intrapersonal influence) or the client (interpersonal influence).

\noindent
\textbf{AIM\_concat}: The model refines the target turn representation by concatenating the GRU output and influence embedding.

% \noindent
% \textbf{AIM\_lin}: A linear layer is used to replace the GRU for encoding the speaker turns. 

\section{Results and Analysis}
\label{sec:res}

\subsection{Performance of the Proposed and Baseline Models}

The classification accuracies of the proposed AIM and the baseline models are presented in Table~\ref{tab:accu_res}. By default, the influence scale $\lambda$ and influence window size $K$ are set to $0.2$ and $3$, respectively. Other values of $\lambda$ and $K$ will be discussed in the following sub-sections.

\begin{table}[htb]
\caption{Classification accuracy on counseling conversations with high vs. low level of therapist empathy}
\label{tab:accu_res}
\centering
\begin{tabular}{c|c}
Model       & Accuracy (\%)      \\ \hline\hline
IM          & 61.1          \\
AIM\_T      & 57.6          \\
AIM\_C      & 62.7          \\
AIM\_concat & 56.8          \\
AIM         & \textbf{69.6}
\end{tabular}
\vspace{-0.05cm}
\end{table}

% AIM\_lin    & 53.4\%          \\

By incorporating both intrapersonal and interpersonal influences, the classification accuracy experiences a notable increase, advancing from 61.1\% to 69.6\%. This improvement provides evidence that our proposed approach, by modeling dynamic influences, significantly facilitates estimating the overall empathy level expressed by the therapist throughout the counseling conversation. The AIM\_C model, which considers only the client's turns within the influence window, slightly outperforms the AIM\_T model, which focuses solely on the therapist's turns. This implies that estimating the therapist's empathy level by analyzing the client's historical behavior may be more effective than analyzing the therapist's own historical behavior. In addition, experimental results indicate that scaled addition is superior to concatenation for incorporating the influence embedding.

\subsection{Impact of Dynamic Influences}

\begin{figure}[htb]
  \centering
  \includegraphics[width=0.8\linewidth]{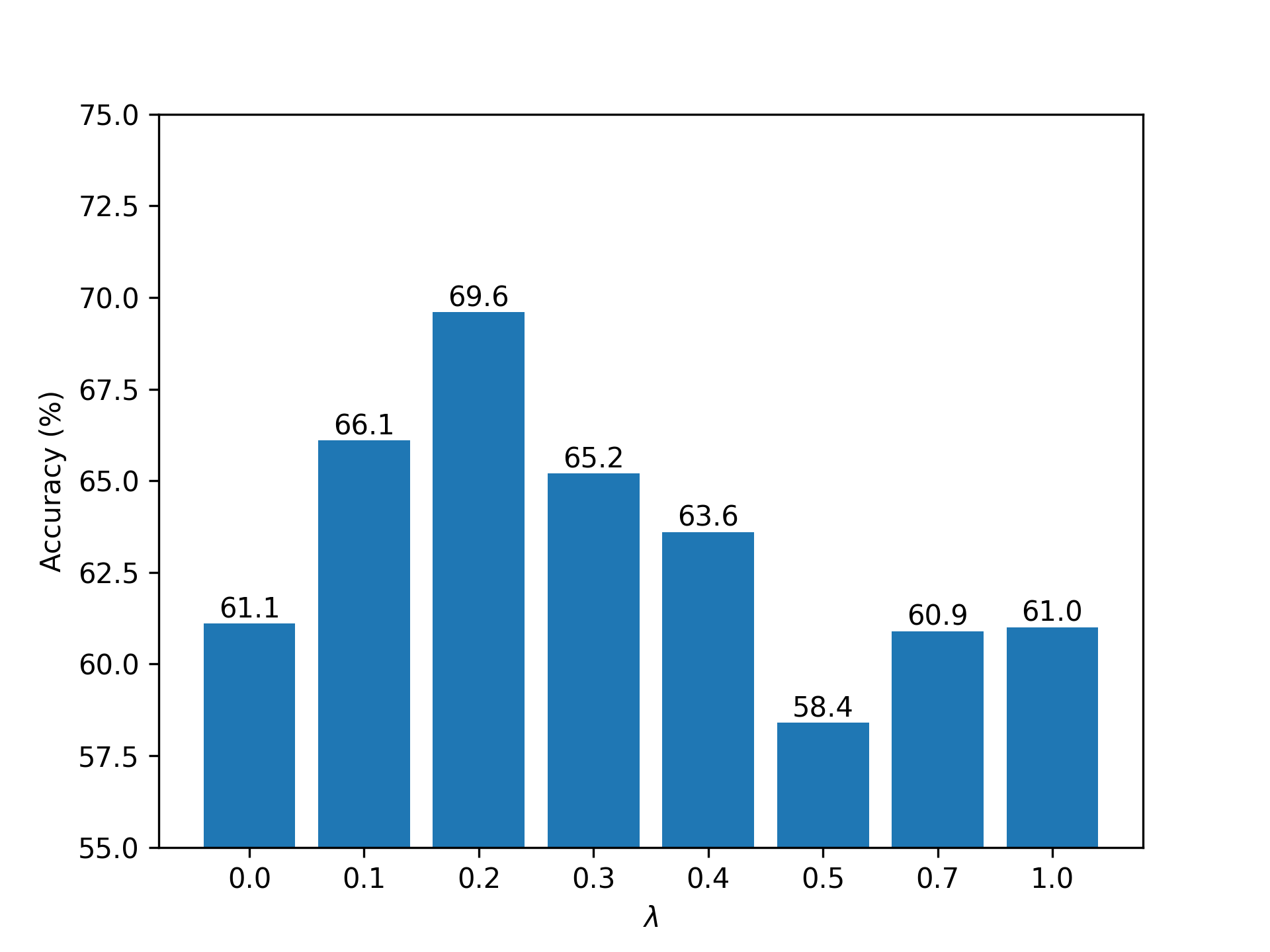}
  \vspace{-0.25cm}
  \caption{Classification accuracy at various values of $\lambda$.}
  \label{fig:attn_scale}
\end{figure}

To quantify the extent to which dynamic influences affect the estimation of therapist empathy, we analyze the performance of the AIM for varying values of the influence scale $\lambda$, as depicted in Figure~\ref{fig:attn_scale}. $\lambda=0.0$ signifies that dynamic influences are not considered in the empathy estimation (equivalent to the baseline model IM). On the other hand, $\lambda=1.0$ suggests that empathy estimation is exclusively dependent on the influences, completely disregarding the therapist's behavior of the current turn. The optimal classification accuracy is achieved at $\lambda=0.2$. This observation suggests that while dynamic influences do contribute to the estimation of therapist empathy, their impact is not excessively dominant.

\vspace{-0.25cm}

\subsection{Optimal number of historical speaker turns }

\begin{figure}[htb]
  \centering
  \includegraphics[width=0.8\linewidth]{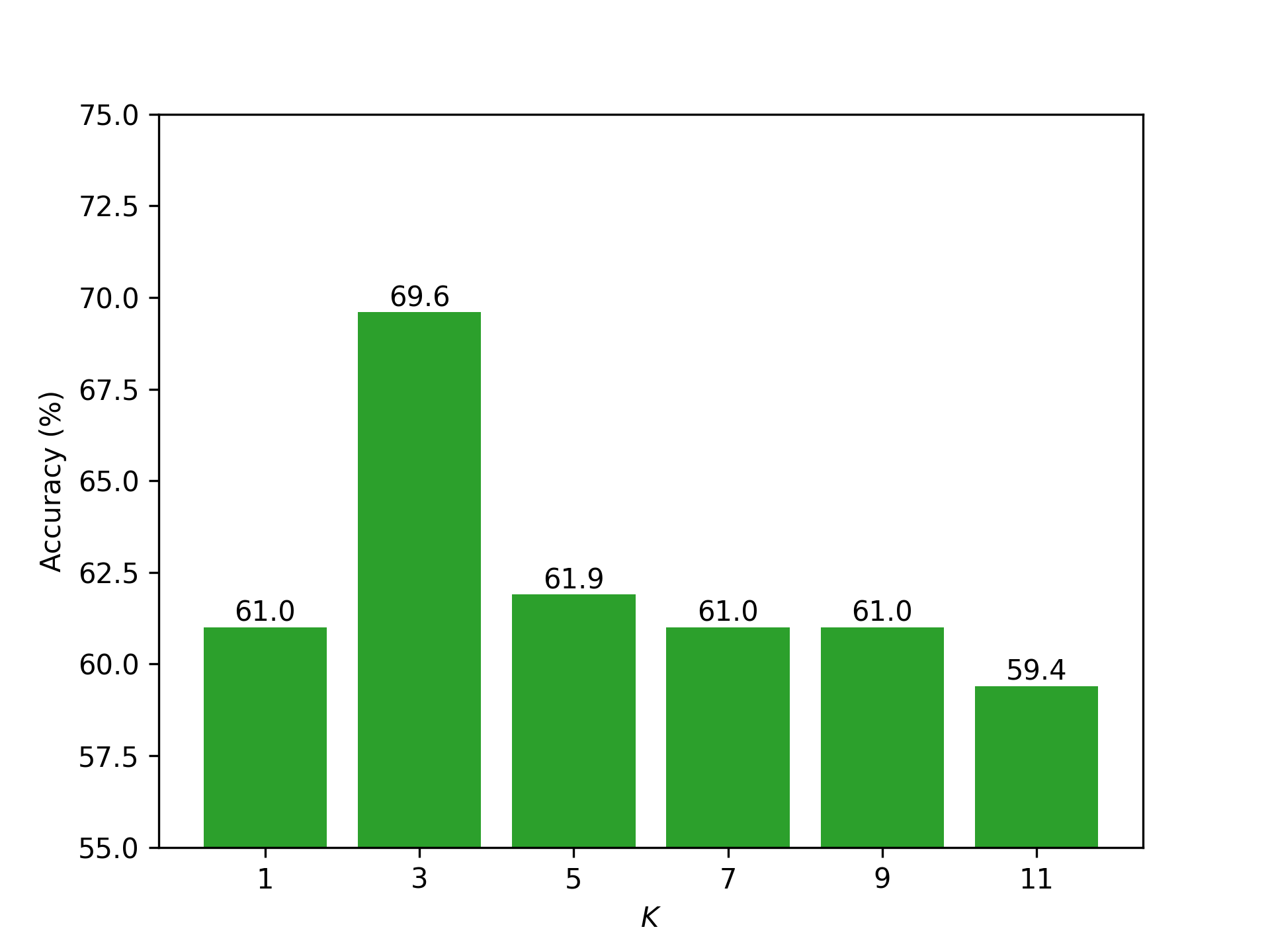}
  \vspace{-0.25cm}
  \caption{Classification accuracy at different values of $K$.}
  \label{fig:influence_win}
\end{figure}

The length of the observation window often plays a crucial role when attempting to assess an individual's behavior through their interaction cues \cite{chakravarthula2021analysis}. In alignment with this understanding, our study seeks to determine the optimal number of historical speaker turns that should be taken into account when estimating therapist empathy. The performance of our proposed model is assessed over a range of influence window sizes, as illustrated in Figure~\ref{fig:influence_win}. The highest classification accuracy is observed at $K=3$. This suggests that focusing exclusively on immediate preceding speaker turns can have a substantial impact on the estimation of therapist empathy within the conversations analyzed.

\vspace{-0.25cm}

\vspace{-0.5cm}
\section{Conclusion}
\label{sec:conc}

In this paper, we propose to use the attention-based influence model to capture both intrapersonal and interpersonal influences in estimating the empathy level of a therapist turn. This is subsequently followed by the median fusion, applied to the turn-level estimates, to determine the therapist's overall empathy level (either high or low) throughout the entire counseling conversation. Our findings indicate that integrating the influence-derived embedding into the target turn's representation improves the estimation of the therapist's empathy level. Notably, the best classification accuracy for empathy level is achieved when this embedding is incorporated in a small proportion. Our study also reveals that estimating the therapist empathy from the client's historical turns is slightly more effective than from the therapist's own historical turns. Additionally, it is observed that focusing solely on the immediate preceding speaker turns can yield an optimal estimation of therapist empathy within analyzed conversations.

\section{Acknowledgements}
\label{sec:ack}
This research is partially supported by the Sustainable Research Fund of the Chinese University of Hong Kong (CUHK) and an ECS grant from the Hong Kong Research Grants Council (Ref.: 24604317).


% \vfill\pagebreak

\bibliographystyle{IEEEbib}
\bibliography{refs}

\end{document}